\documentclass[usenatbib]{mnras}
\usepackage{newtxtext,newtxmath}
\usepackage[T1]{fontenc}

\DeclareRobustCommand{\VAN}[3]{#2}
\let\VANthebibliography\thebibliography
\def\thebibliography{\DeclareRobustCommand{\VAN}[3]{##3}\VANthebibliography}

\usepackage{CJK}
\usepackage{amsmath}
\usepackage{graphicx}   % need for figures
\usepackage{color}      % use if color is used in text
\usepackage{soul}
\usepackage{float}

\usepackage{hyperref}
    \newcommand{\vct}[1] {\boldsymbol{#1}}
  \renewcommand{\d}{\mathrm{d}}
    \newcommand{\e}{\mathrm{e}}

   \newcommand{\di} {d_\text{p}}
   \newcommand{\de} {d_\text{e}}
   \newcommand{\Rey} {\mathrm{Re}}
   \newcommand{\PiD} {\text{Pi-D}}

\title{Effective Viscosity, Resistivity, and Reynolds Number in Weakly Collisional Plasma Turbulence}
\author[Y. Yang et al.]{
Yan Yang,$^{1}$\thanks{E-mail: yanyang@udel.edu}
William H. Matthaeus,$^{1}$
Sean Oughton,$^{2}$
Riddhi Bandyopadhyay,$^{3}$
Francesco Pecora,$^{1}$
\newauthor{
Tulasi N. Parashar,$^{4}$
Vadim Roytershteyn,$^{5}$
Alexandros Chasapis,$^{6}$
and Michael A. Shay$^{1}$}
\\
$^{1}$Department of Physics and Astronomy, University of Delaware, Newark, DE 19716, USA\\
$^{2}$Department of Mathematics, University of Waikato, Hamilton 3240, New Zealand\\
$^{3}$Department of Astrophysical Sciences, Princeton University, Princeton, NJ 08544, USA\\
$^{4}$School of Chemical and Physical Sciences, Victoria University of Wellington, Wellington 6012, New Zealand\\
$^{5}$Space Science Institute, Boulder, CO 80301, USA\\
$^{6}$Laboratory for Atmospheric and Space Physics, University of Colorado Boulder, Boulder, CO 80309, USA
}

\begin{document}
\maketitle

\begin{abstract}
We examine dissipation and energy conversion in weakly collisional plasma turbulence, employing \emph{in situ} observations from the Magnetospheric Multiscale (MMS) mission and kinetic Particle-in-Cell (PIC) simulations of proton-electron plasma.
%% We employ kinetic Particle-in-Cell simulations of proton-electron plasma as well as \emph{in situ} observations from the Magnetospheric Multiscale mission, to examine dissipation and energy conversion in weakly collisional turbulence.
A previous result indicated the presence of viscous-like and resistive-like scaling of average energy conversion rates---analogous to scalings characteristic of collisional systems.  This allows for extraction of collisional-like coefficients of \emph{effective} viscosity and resistivity, and thus also 
%%  direct 
determination of effective Reynolds numbers based on these coefficients. 
The effective Reynolds number, as a measure of the available bandwidth for turbulence to populate various scales, links macro turbulence properties with kinetic plasma properties in a novel way. 
\end{abstract}

\section{Introduction}

%---------------------------------------------  

Energy dissipation in fluids and plasmas may be effectively defined as the conversion process by which 
macroscopic reservoirs of 
energy are transformed into heat. 
Mechanisms of
energy dissipation for weakly collisional or collisionless plasma are of central importance for addressing long-standing fundamental problems in space and astrophysics.  These include, for example, 
the acceleration of energetic particles and the heating of the solar corona and solar wind.
In collisional cases, the  (viscous and resistive) dissipation is expressed in a simple form in terms of viscosity, resistivity, and spatial gradients of the velocity and magnetic fields. 
However, space plasmas frequently 
reside in (nearly) collisionless regimes, where the dissipation mechanisms are not well understood.
For example, in one of the most well-studied space plasmas, the solar wind \citep{BrunoCarboneLRSP13}, the collision length is of the order of 1\,AU and collisions are typically too weak to establish a local equilibrium (Maxwellian particle distribution) \citep{Marsch06,verniero2020parker}. In such cases the classical collisional approach becomes generally 
inapplicable, as do standard closures that describe dissipation in terms of fluid-scale variables and viscosity and resistivity.

Lacking the standard collisional closures, 
studies of plasma turbulence 
have shown increasing interest in quantifying collisionless dissipation. 
Investigations of collisionless dissipation have often considered one or more of the following three aspects: 

(i) \emph{Dissipation mechanisms.} 
Collisionless dissipation has often been described in terms of specific mechanisms such as magnetic reconnection \citep{RetinoEA07}, 
wave-particle interaction \citep{MarkovskiiEA06,HowesEA08-prl,ChandranEA10-perpHeat},
and turbulence-driven intermittency \citep{DmitrukEA04-tp,ParasharEA11}. 
Identification of such processes affords specific physical insight. If all possible mechanisms and their relative contributions can be identified, a full understanding of the dissipation physics may be achievable. 

(ii) \emph{Turbulence cascade.} 
The picture of turbulence cascade describes energy transfer across scales from an energy-containing range, through an inertial range, and into a (small-scale) dissipation range. Different dissipation proxies based on the turbulence cascade process have been adopted to estimate the dissipation rate. At energy-containing scales, the global decay rate of energy is controlled by the von K\'arm\'an decay law 
  \citep{KarmanHowarth38,HossainEA95,WanEA12-vKH,ZankEA17-NIxport}. 
At inertial range scales, the Yaglom relation 
\citep{PolitanoPouquet98-grl,Sorriso-ValvoEA07,HadidEA17,andres2019energy,banerjee2020scale} 
has been adapted to estimate the energy transfer rate. 
\citet{hellinger2022ion} extended this approach into the kinetic range by empirically including pressure-strain interaction effects in the kinetic range. 

(iii) \emph{Energy conversion channels.} 
Yet another approach to understand dissipation is to trace the flow of energy and examine energy conversion between different forms. Temperature enhancement implies increase of thermal energy and to specifically track  thermal energy production requires quantification of energy supplies from energy reservoirs for each species. 
Two widely-invoked classes of 
  %% energy 
conversion are the electric work on particles for species $\alpha$, $\vct{J}_{\alpha} \cdot \vct{E}$ \citep{ZenitaniEA11} and the pressure-strain interaction for species $\alpha$, $-\left( \vct{P}_{\alpha} \cdot \nabla \right) \cdot \vct{u}_{\alpha}$ 
  \citep{YangEA17-PoP,YangEA17-PRE}.
(We employ a familiar plasma physics notation with full definitions given in
  Sec.~\ref{sec:theory}.)
These channels play different roles:
the electric work measures the release of electromagnetic energy, while the pressure-strain interaction measures the increase of thermal energy.

Collisional and collisionless dissipation obviously differ from each other, but they also share similarities. For example, in both cases, conversion of energy between different forms can be quantified in terms of pressure work and electric work. 
  %% in both collisional and collisionless cases. 
In collisional cases, however, these two channels can be further approximated as viscous dissipation via velocity gradients and resistive dissipation via electric current density (i.e, magnetic field gradients), which will be discussed in detail in Sec.~\ref{sec:theory}. 
On the other hand, investigations using \emph{in situ} spacecraft data \citep{ChasapisEA18-diss,BandyopadhyayEA20-PiD} and numerical simulations \citep{WanEA16,YangEA17-PoP} support a novel and less obvious 
idea, namely that  
collision\emph{less} dissipation 
is also in direct association with velocity strain rate and electric current density \citep{BandyopadhyayEA2023-collisionallike}.  More specifically, by quantifying collisionless dissipation by the electric work $\vct{J}_{\alpha} \cdot \vct{E}$ and the pressure-strain interaction $-\left( \vct{P}_{\alpha} \cdot \nabla \right) \cdot \vct{u}_{\alpha}$, these are seen to be well correlated with, respectively, squared electric current density and squared velocity strain rate.  
This association stands in direct analogy to the resistive and viscous dissipation in collisional plasmas. 
It is natural then to inquire more deeply 
into the behavior of
collisionless dissipation and its 
similarities with collisional dissipation. 

Initial steps in this direction have shown two findings.
First, 
the global average of electric work  conditioned on electric current density scales as $J^2$, i.e., the square of the current density
  \citep{WanEA16,ChasapisEA18-diss}.
Second, that there is a 
similar scaling of pressure work with respect to $D^2 = D_{ij} D_{ij}$, where $D_{ij}$ is the traceless velocity strain rate tensor
  \citep{BandyopadhyayEA2023-collisionallike}. 
These results provide strong evidence supporting the concept of collisional-\emph{like} dissipation in collisionless plasmas, and, moreover, allow a novel estimation of effective viscosity and resistivity, which is then further applied to define \emph{effective Reynolds numbers}.

Since the classical closures of viscosity and resistivity are inapplicable to collisionless plasmas, one might suspect that various features of classical turbulence theory might not 
be applicable, in particular regarding dissipative processes and the several length scales and dimensionless numbers related to dissipation.  
Even the notion of Reynolds number ($ \Rey $)---which in the hydrodynamic sense is the ratio of the strengths of nonlinear and viscous effects---needs to be considered with caution in the absence of viscosity and resistivity.
On the other hand a point of encouragement is that wavenumber spectra in large collisionless 
plasmas such as the solar wind \citep{BrunoCarboneLRSP13}
often exhibit a Kolmogorov-like power-law energy spectrum \citep{Coleman68}
that extends from a correlation scale \citep{MattEA05-corrn}
to smaller kinetic scales \citep{LeamonEA98-jgr}, below which the spectrum steepens. 
Between these scales the power-law inertial range is expected to span a larger range when the Reynolds number  is greater, by analogy with hydrodynamics. 

  %% But $ \Rey $, classically the ratio of the strengths of nonlinear and viscous effects, clearly requires reconsideration in the collisionless case.

To achieve physically motivated 
generalizations of $ \Rey $ in the collisonless case, 
previous studies have adopted various definitions of effective Reynolds number, often related to the ratio of an outer scale to an inner scale. For example
\begin{equation}
   \Rey \approx \left( \frac{\lambda_c} { \lambda_d} \right)^{4/3}
   \quad \text{or} \quad
   \Rey \approx \left( \frac{\lambda_c} {\lambda_T} \right)^2,
 \label{eq:Reff-inner}
\end{equation}
where $\lambda_c$ is the correlation length, $\lambda_d$ is a dissipation scale, 
and $\lambda_T$ is the Taylor microscale
   \citep{BatchelorTHT,Pope}. 
For a weakly collisional plasma, such as the solar wind, the dissipation scale can be presumed to be
the ion inertial length $\di$ or the ion thermal gyroradius \citep{Verma96a,ParasharEA19,CuestaEA22-intermit}, given that the inertial-range spectrum terminates 
(and then steepens) 
near these scales
\citep{LeamonEA98-jgr,SmithEA06-diss,MattEA08-Taylor,ChenEA14-grl}. 
Another scale related to dissipation is the Taylor microscale $\lambda_T$.
This has been measured in the solar wind and then used to estimate effective Reynolds number for that system
  \citep{MattEA05-corrn,MattEA08-Taylor,ChuychaiEA14, BandyopadhyayEA20-Taylor,PhillipsEA22-lengths}.
Note that both of the empirical determinations of effective Reynolds number given by 
    Eq.~\eqref{eq:Reff-inner} 
depend on the appropriate estimates of inner scales. 
Herein we adopt a different approach that avoids any need to estimate inner scales.
In a novel examination of the putative connection between collisional and collisionless dissipation, 
we explore specific evaluations of effective viscosity, resistivity, and Reynolds number from 2.5D and 3D kinetic Particle-in-Cell (PIC) simulations and \emph{in situ} observations from the \emph{Magnetosphere Multiscale} (MMS) mission. 

\section{Theoretical Background}
\label{sec:theory}
We are concerned with observed phenomena related to energy conversion processes and focus on the bulk flow energy, electromagnetic energy, thermal energy,
and the conversion and dissipation channels that link them. 
For consistency in the contexts of 
observational and simulation
data 
all quantities will be expressed in SI units throughout the paper.

    \subsection{ Collisional cases}
    \label{sec:collisonal}
    
We start with the simplest one-fluid magnetohydrodynamic (MHD) model. The momentum and magnetic induction equations read,
\begin{eqnarray}
\rho \frac{\partial \vct{u}}{\partial t} + \rho \vct{u} \cdot \nabla \vct{u} &=& -\nabla p -\nabla \cdot \vct{\Pi} + \vct{J} \times \vct{B}, \label{eq:mhd-u}\\
\frac{\partial \vct{B}}{\partial t}-\nabla \times (\vct{u}\times \vct{B}) &=& \eta \nabla^2 \vct{B}, \label{eq:mhd-b}
\end{eqnarray}
where $\Pi_{ij} = -\mu \left( \partial_i u_j + \partial_j u_i \right)
         + {\frac{2}{3}} \mu (\nabla \cdot  \vct{u}) \delta_{ij}$ 
is the viscous stress tensor, $\vct{J}=\frac{1}{\mu_0}\nabla \times \vct{B}$ is the electric current, $\mu$ is the dynamic viscosity, $\eta$ is the magnetic diffusivity, and $\mu_0 = 4\pi \times 10^{-7}~[\rm{H\cdot m^{-1}}]$ is the magnetic permeability of free space
(aka vacuum permeability).
%To make a direct comparison between PIC simulations and MMS intervals, all quantities will be expressed in SI units throughout the paper.

Based on Eqs.~\eqref{eq:mhd-u} and \eqref{eq:mhd-b} 
one readily obtains the collisional dissipation rates of bulk flow energy density 
($E_u=\frac12\rho \vct{u}^2$) 
and magnetic energy density 
($E_b=\frac{1}{2\mu_0} \vct{B}^2$).
These can be
expressed in terms of the coefficients of 
dynamic viscosity ($\mu$) 
and electrical resistivity ($1/\sigma \equiv \mu_0 \eta$), 
and particular pieces of the velocity gradient and magnetic gradient tensors:
\begin{eqnarray}
    D_{\mu} &=& 2\mu D^2,\label{eq:vis_diss}
 \\
    D_{\eta} &=& \frac{1}{\sigma} J^2. 
 \label{eq:res_diss}
\end{eqnarray}
Here 
  $ D_{ij} = \frac{1}{2} ( \partial_i u_j + \partial_j u_i)- \frac13(\nabla \cdot \vct{u}) \delta_{ij}$ 
is the traceless strain rate tensor, 
with $D^2 = D_{ij}D_{ij} $ the second invariant of $D_{ij}$
and equal to the sum of the squares of the eigenvalues of $D_{ij}$.

The viscous dissipation in Eq.~\eqref{eq:vis_diss} and the resistive dissipation in Eq.~\eqref{eq:res_diss} are actually the 
    \emph{closures} 
of the anisotropic part of 
$-\left( \vct{P} \cdot \nabla \right) \cdot \vct{u}$ 
and the electric work 
$\vct{J} \cdot \vct{E}$ 
(see Eqs.~\eqref{eq:Ef}--\eqref{eq:Em}) 
in the presence of frequent collisions. 
These can be derived by kinetic methods \citep{chapman1939mathematical,marshall1960kinetic,Braginskii65,kaufman1960plasma}, where an approximate solution for the Boltzmann equation is firstly obtained in terms of macroscopic variables (like density, velocity, and temperature) and the pressure tensor (the second-order moment of the velocity distribution function) is then also expressed in terms of macroscopic variables. 
The closure of the electric work (i.e., Eq.~\eqref{eq:res_diss}) can be derived using Ohm's law. The detailed procedure is: 
(i) In the presence of frequent collisions, it can be shown that no matter what the initial conditions are the velocity distribution function (VDF) $f$ must approach a Maxwellian $f_0$ in a time of the order of the mean time between collisions \citep{chapman1939mathematical}. 
(ii) The VDF $f$ is assumed to be approximately a Maxwellian $f_0$ and high-order terms ($f_1$, $f_2$, $\cdots$) are introduced as small corrections or perturbations on the Maxwellian distribution function, $f=f_0+f_1+f_2+\cdots$.
%\begin{equation}
%    f=f_0+f_1+f_2+\cdots \label{eq:f_expansion}
%\end{equation}
(iii) Retaining only the first-order correction $f_1$ and disregarding higher-order terms can give rise to the collisional dissipation, i.e., Eqs.~\eqref{eq:vis_diss} and \eqref{eq:res_diss}.

%The Coulomb collisions (not true collisions) in plasma are the interaction between charged particles due to the long range of the Coulomb force. But the real collision terms are extremely complicated, which are always described in a simplified form, such as the so-called Boltzmann collision operator. 
To prescribe the 
applicability of the collisional approximation, we should keep in mind its requirement:
\emph{Even though the collisional dissipation provides a simple representation of dissipation in terms of the viscosity and resistivity, 
in all standard cases it applies only when the local distribution is very close to a Maxwellian due to particle collisions.}

    \subsection{ Collisionless cases}
    \label{sec:collisionless}
    
The time evolution of energies can be derived using the first three moments of the Boltzmann equation, in conjunction with the Maxwell equations.  One obtains \citep{Braginskii65,ChiuderiVelli,YangEA17-PoP,YangEA17-PRE}
\begin{eqnarray}
\partial_t \mathcal{E}^{f}_\alpha + \nabla \cdot \left( \mathcal{E}^{f}_\alpha \vct{u}_\alpha +  \vct{P}_\alpha \cdot \vct{u}_\alpha \right) &=& \left( \vct{P}_\alpha \cdot \nabla \right) \cdot \vct{u}_\alpha+ \vct{J}_\alpha \cdot \vct{E}, \label{eq:Ef}\\
\partial_t \mathcal{E}^{th}_\alpha + \nabla \cdot \left( \mathcal{E}^{th}_\alpha \vct{u}_\alpha +\vct{h}_\alpha\right) &=& -\left( \vct{P}_\alpha \cdot \nabla \right) \cdot \vct{u}_\alpha, \label{eq:Eth}\\
\partial_t \mathcal{E}^{m} +  \nabla \cdot \left( \vct{E} \times \frac{\vct{B}}{\mu_{0}} \right) &=& -\vct{J} \cdot \vct{E}, \label{eq:Em}
\end{eqnarray}
where the subscript $\alpha=e, p$ represents the particle species (electrons and protons).
Here,
$\mathcal{E}^m = \dfrac{1}{2}\left(\epsilon_0 \vct{E}^2+\vct{B}^2/\mu_{0}\right)$
is the electromagnetic energy density, 
with $\vct{E}, \vct{B}$ the electric and magnetic fields;
$\mathcal{E}^f_\alpha = {\frac{1}{2}}\rho_\alpha \vct{u}^2_\alpha$
is the bulk flow energy density for species $\alpha$, 
with mass density $\rho_\alpha$ 
and bulk flow velocity $\vct{u}_\alpha$;  
$\mathcal{E}^{th}_\alpha = {\frac{1}{2}} m_\alpha \int_v{(\vct{v}-\vct{u}_\alpha) \cdot (\vct{v}-\vct{u}_\alpha) f_\alpha \, \d^3{v}} $ 
is the thermal energy,
with mass $m_\alpha$ and velocity distribution function $f_\alpha( \vct{x}, \vct{v} )$;
$\vct{P}_\alpha$ is the pressure tensor;
$\vct{h}_\alpha$ is the heat flux vector;
$\vct{J}=\sum_{\alpha} \vct{J}_\alpha$ is the total electric current density with
$\vct{J}_\alpha = n_\alpha q_\alpha \vct{u}_\alpha$ the electric current density of species $\alpha$; 
$n_\alpha(\vct{x})$ and $q_\alpha$ are the number density and the charge of species $\alpha$, respectively. 
As we can see the energy conversion between bulk flow and thermal is quantified by the pressure-strain interaction,
  $-\left( \vct{P}_{\alpha} \cdot \nabla \right) \cdot \vct{u}_{\alpha}$, 
while the energy conversion between bulk flow and electromagnetic is quantified by the electric work, $\vct{J} \cdot \vct{E}$.
We emphasize that there are no $\vct{J}_\alpha$ terms in the thermal energy equation~\eqref{eq:Eth}.

The basic assumption of collisional dissipation is that inter-particle collisions are sufficiently strong to maintain a local equilibrium. In principle, this assumption is not valid in collisionless plasmas.
Instead, the particle VDF often displays a distorted out-of-equilibrium shape characterized by non-Maxwellian features as observed in \emph{in situ} data    
   \citep{graham2017instability,perri2020deviation} 
and in numerical simulations   \citep{ServidioEA12}.
Although collisionless plasma can be described by the (collisional) MHD model at large scales, spacecraft \emph{in situ} measurements reveal complex features at kinetic scales. At these small scales, kinetic processes must take place. One widely accepted picture of solar wind fluctuations is that they are characterised by broadband energy spectra with several spectral breaks and spectral steepening at kinetic scales 
  \citep{LeamonEA98-jgr,SahraouiEA09,AlexandrovaEA09,KiyaniEA15}. 
In particular, observations indicate that the steepening of velocity and magnetic field spectra at kinetic scales is clearly dependent on the dissipation rate 
    \citep{SmithEA06-diss}. 
Clearly collisionless dissipation delves deeply into kinetic plasma processes. 
Unlike collisional dissipation terminating at dissipation scales, collisionless dissipation is dominant at 
a range of kinetic scales.

    \subsection{Similarities between collisional and collisionless dissipation} 
    \label{sec:similarities}
Even though collisionless dissipation differs from collisional dissipation in several ways,  studies also suggest that there are similarities between them. 
First, they are both organized in structured patterns and concentrated at, or near, coherent structures. Coherent structures form dynamically in MHD and plasma flows and are found to be of importance in heating.
They include current sheets and vortices.
According to the definition of collisional dissipation in Eqs.~\eqref{eq:vis_diss} and \eqref{eq:res_diss}, it should not be at all surprising to find that the collisional dissipation occurs with intense values at (and near) these structures. 
The physical quantities that are responsible for the conversion of energy in collisionless plasmas 
(see Eqs.~\eqref{eq:Ef}-\eqref{eq:Em}) 
are also found in the same kind of spatial localization 
  \citep{OsmanEA11-swHeat,RetinoEA07,YangEA17-PoP,ServidioEA12,FranciEA16,ParasharEA16-om}.
In this sense, both collisional and collisionless dissipation concentrates in structured patterns. 
Second, they are both in direct association with velocity strain rate and electric current density. As we have already remarked, collisionless dissipation, as quantified by the electric work 
  $\vct{J}_{\alpha} \cdot \vct{E}$ 
and the pressure-strain interaction 
  $-\left( \vct{P}_{\alpha} \cdot \nabla \right) \cdot \vct{u}_{\alpha}$, 
is found to be in direct association with
    $J^2$ and $D^2$
 \citep{ChasapisEA18-diss,BandyopadhyayEA20-PiD,WanEA16,YangEA17-PoP,BandyopadhyayEA2023-collisionallike},
and this scaling is analogous to the resistive and  viscous dissipation that 
Eqs.~\eqref{eq:vis_diss} and \eqref{eq:res_diss} represent.

We therefore conjecture that a closure for collisionless dissipation that is similar to collisional dissipation is plausible,
in a statistical sense, the details of which are to be determined. 
That is, we suggest that
\begin{eqnarray}
    \langle -\Pi_{ij}D_{ij} |D\rangle 
    & \sim& 2\mu D^2, \label{eq:vis-scaling} \\
    \langle {\vct J} \cdot {\vct E}'|J \rangle 
    & \sim & {\frac{1}{\sigma}} J^2, \label{eq:res-scaling}
\end{eqnarray}
where $\Pi_{ij} = P_{ij} - p\delta_{ij}$ 
is the deviatoric pressure tensor, and 
$ \vct{E}' = \vct{E} + \vct{u}_e \times \vct{B}$
is the electric field in the electron fluid frame. 
$\langle -\Pi_{ij}D_{ij} |D\rangle$ is the average  
of the anisotropic part of the pressure-strain
interaction conditioned on 
 $D \equiv \sqrt{ D_{ij}D_{ji} } $, and
$\langle \vct{J} \cdot \vct{E}^\prime |J \rangle$ is the average
of the electric work in the electron fluid frame conditioned on the (local) current magnitude 
    $J = |\vct{J} |$.
If the scalings in Eqs.~\eqref{eq:vis-scaling} and \eqref{eq:res-scaling}
can be verified, 
they will permit an evaluation of effective values for dynamic viscosity $\mu$ and electrical resistivity $1/\sigma$, 
and thence for effective 
kinematic viscosity   $\nu  = {\mu}/{\rho}$
and
magnetic diffusivity  $\eta = 1/(\sigma \mu_0)$ .

    \section{Data}
    \label{sec:data}

We present data from 2.5D and 3D 
fully kinetic PIC simulations and one long MMS burst-mode interval in the magnetosheath. In each case the analysis leads to a determination of 
the associated (effective) resistivity and separate viscosities for electrons and protons.

The \emph{2.5D PIC simulation} 
employs the P3D code \citep{ZeilerJGR02}, which has also been 
used in \citet{YangEA22,YangEA23-agyro} 
and \citet{BandyopadhyayEA2023-collisionallike}.
Here 2.5D means, as usual, that there are
three components of dependent
field vectors and a 2D spatial grid,
i.e., that the phase space coordinates are $ (x,y, v_x,v_y,v_z) $.
Normalization in P3D is largely ``proton-based'',  
with 
number density normalized to a reference value $n_r$,
mass to proton mass $m_\text{p}$,
charge to proton charge $e$, 
and
magnetic field to a reference $B_r$.
Length is normalized to the proton inertial length $\di$,
time to the proton cyclotron time $\omega_{cp}^{-1}$,
and velocity to the 
  consequent
reference Alfv{\'e}n speed
$V_{Ar} = B_r/\left(\mu_0 m_\text{p} n_r\right)^{1/2}$.

The particular simulation we consider is performed in a square periodic domain 
of size $L=150 \, \di$, with $4096^2$ grid points and
$3200$ particles of each species
per cell ($\sim 1.07\times 10^{11}$ total particles).
For numerical expediency we employ artifically low values of
the proton to electron mass ratio, $m_\text{p}/m_\text{e} = 25$,
and
the speed of light, $ c=15 \,V_{Ar} $.
The run is
a decaying initial value problem,
starting with uniform densities
and temperatures for both species.
A uniform magnetic field, $B_0 = 1.0$, is directed
out of the plane, and the initial plasma $\beta$ is 0.6.
%The initial $\vct{v}$ and $\vct{b}$ fluctuations are transverse to $B_0$
%(``Alfv\'en mode'') and have 
%Fourier wavevectors $2\le |\vct{k}L/(2\pi)| \le 4$. The Fourier modes have
%random phases. 
%with amplitudes chosen to yield 
%spectra proportional to $ \left[ 1 + \left(k/k_0\right)^{8/3} \right]^{-1} $, with  $ k_0 = 6 $. 
The initial $\vct{v}$ and $\vct{b}$ fluctuations are transverse to $B_0$
(``Alfv\'en mode'') and have 
Fourier modes with random phases for the wavenumber range $2\le |\vct{k}L/(2\pi)| \le 4$. 
The initial normalized cross helicity $\sigma_c$ is negligible.

The \emph{3D simulation} \citep{RoytershteynEA15} is 
obtained using the VPIC code \citep{BowersEA08}, which was also used in \citet{YangEA22}
and \citet{BandyopadhyayEA2023-collisionallike}.
VPIC normalization differs significantly 
from that in P3D, being more electron based. 
Number density is normalized to a reference value $n_r$,
mass to electron mass $m_\text{e}$,
charge to proton charge $e$,
length to the electron inertial length $\de$,
time to the electron plasma oscillation time $\omega_{pe}^{-1}$,
velocity to the 
  (true physical)
speed of light $c$,
and magnetic field to a reference $B_r = m_\text{e} c \omega_{pe}/e$.

The simulation of interest herein was performed in a cubic 
periodic domain of size
$L=296 \, \de$, with $2048^3$ grid points and
$150$ particles of each species
per cell ($\sim 2.6\times 10^{12}$ total particles).
The proton to electron mass ratio is $m_\text{p}/m_\text{e} = 50$.
Like the P3D run, this one is also
a decaying initial value problem,
starting with uniform density
and temperature of protons and electrons.
There is a uniform magnetic field $B_0 = 0.5$ in the out-of-plane
  $ \hat{\vct{ z}} $ direction,
%% directed out of the plane, 
and the plasma $\beta$ is 0.5.
The $\vct{v}$ and $\vct{b}$ fluctuations are initialized with
two orthogonal polarizations and an overall power spectrum decaying as $k^{-1}$ for the wavenumber range $1\le |\vct{k}L/(2\pi)| \le 7$ with equal power in each polarization.
The initial $\vct{v}$ and $\vct{b}$ fluctuations are a mixture of Alfv\'enic and randomly phased perturbations. The initial normalized cross helicity is $\sigma_c \simeq 0.44$. 

Key parameters for the 2.5D and 3D runs are given in
Table~\ref{tab:2.5d-3d-params}.
For both runs we analyze statistics
at a time shortly after that at which the 
maximum mean square current density occurs.
Prior to analyses, we remove noise inherent in the PIC plasma algorithm via a
low-pass Fourier filtering of the fields.
\begin{table}   %[H]
    \centering
    \begin{tabular}{cccccccc}
    \hline
	    Dimension  & $L$ & $N$ & $m_\text{p}/m_\text{e}$ & $B_0 \hat{\vct{z}}$  & $\delta b/B_0$ & $\beta$ & \emph{ppg} \\
     \hline
	    2.5D   & $150 \, \di$ & 4096 & 25 & 1.0 & 0.5 & 0.6 & 3200 \\
     \hline
          3D    & $296 \, \de$ & 2048 & 50 & 0.5 & 1.0 & 0.5 & 150 \\
     \hline
    \end{tabular}
    \caption{2.5D and 3D PIC simulation parameters in code units: domain size $L$; 
    grid points in each direction $N$; 
    proton-to-electron mass ratio $m_\text{p}/m_\text{e}$; 
    guide magnetic field in  $z$-direction $B_0$; initial magnetic fluctuation amplitude $\delta b$; 
    plasma $\beta$; 
    average number of particles of each species per grid \emph{ppg}.}
    \label{tab:2.5d-3d-params}
\end{table}

In addition to simulation data we also analyze an interval of MMS spacecraft data. 
The MMS mission provides high time cadence and simultaneous multi-spacecraft measurements, typically in a tetrahedral formation, with small inter-spacecraft separations.
The MMS spacecraft sample the near-Earth plasma including the magnetosheath \citep{BurchEA16}. The proton and electron three-dimensional velocity distribution functions (VDFs) are available from the Fast Plasma Investigation (FPI) \citep{PollockEA16} instrument. One can then determine density, velocity, 
pressure tensor, and current density, with a time resolution of 
   150\,ms
for ions and 
    30\,ms 
for electrons.
The Flux-Gate Magnetometer (FGM) \citep{Russell2016} measures the vector magnetic field, and the Electric Field Double Probes (EDP) \citep{ErgunEA16-SSR} measures the electric field.
Herein we employ a single long-burst interval of MMS data obtained in the magnetosheath 
 %% from 06:12:43 to 06:52:23 on 2017 December 26; 
(see Table~\ref{tab:MMS}).
For this interval the mean plasma velocity is approximately 
   230\,km\,s$^{-1} $
   and
the inter-spacecraft separation is about 
  27\,km, 
which is below the ion inertial length and corresponds to a few times the electron inertial length. 
As shown in previous studies \citep{ParasharEA18-prl,BandyopadhyayEA20-PiD,YangEA23-agyro,BandyopadhyayEA2023-collisionallike}, 
this interval exhibits features of well-developed turbulence.
%These conditions enable reliable evaluation of the velocity strain tensor, employing techniques analogous to the curlometer technique \citep{DunlopEA02}. 
\begin{table*} 
    \centering
    \begin{tabular}{rrccccccc}
    \hline
	   & $|\langle \vct{B}\rangle|$[nT]  & $\delta B/|\langle \vct{B}\rangle|$ & $\langle n_\text{e}\rangle[\rm{cm}^{-3}]$ & $\langle n_\text{p}\rangle[\rm{cm}^{-3}]$ & $\beta_p$ & $\di$[km]  &$\de$[km]  & $L$[km] \\
     \hline
      2017 Dec 26 06:12:43-06:52:23 & 22.0  & 0.8 & 24.9 & 22.8 & 4.5 & 48 & 1.1 & 27 \\
     \hline
     \label{tab:MMS}
    \end{tabular}
    \caption{Description of one selected magnetosheath interval of MMS data. 
    $ | \langle \vct{B} \rangle | $ is the mean magnetic field strength;
    $\delta B = \sqrt{\langle |\vct{B}(t)-\langle \vct{B}\rangle|^2\rangle}$ is the root-mean-square magnetic fluctuation; 
    $ \langle n \rangle $ is the mean plasma density; 
    $ \beta_p $ is the proton plasma beta; 
    $ \di $ and $ \de $ the ion and electron inertial lengths; 
    $ L $ indicates the mean separation between spacecraft.
    }
\end{table*}

    \section{Results}
    \label{sc:results}
    
    \subsection{Kinematic Viscosity and Magnetic Diffusivity}
    \label{sec:res-diff}

\begin{figure*}
    \centering
    2.5D P3D \hspace{18em} 3D VPIC \hspace{18em} MMS\\
    \includegraphics[width=0.32\textwidth]{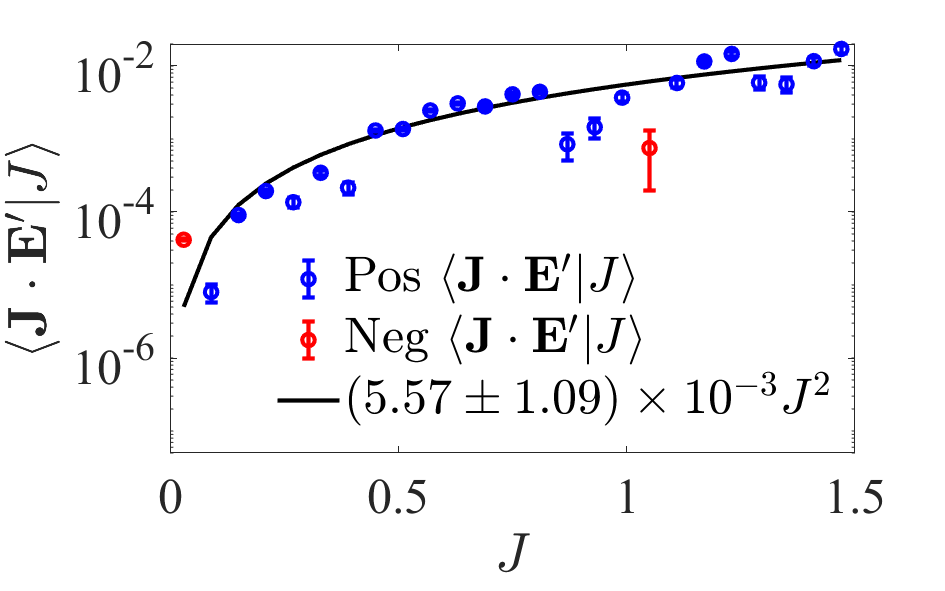}
    \includegraphics[width=0.32\textwidth]{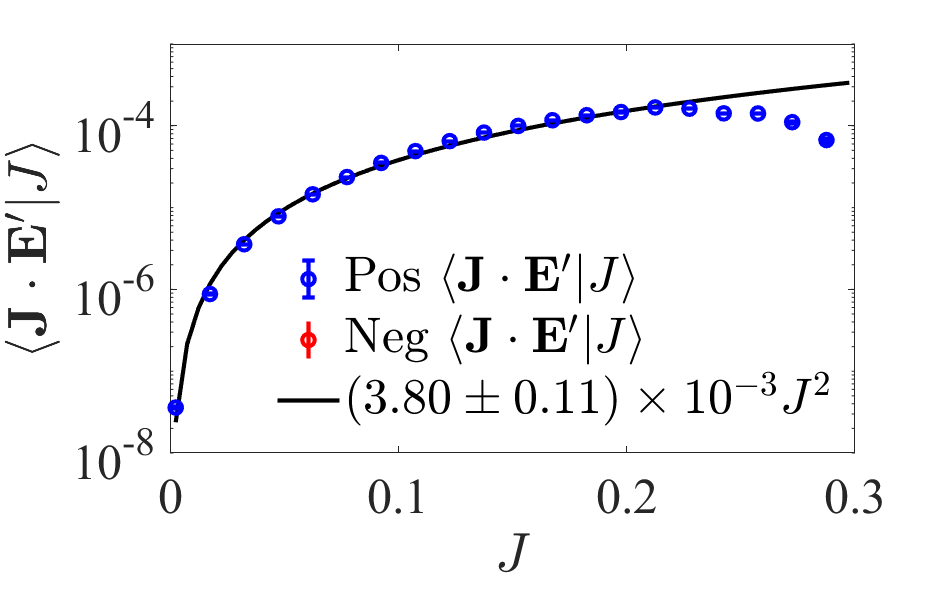}
    \includegraphics[width=0.32\textwidth]{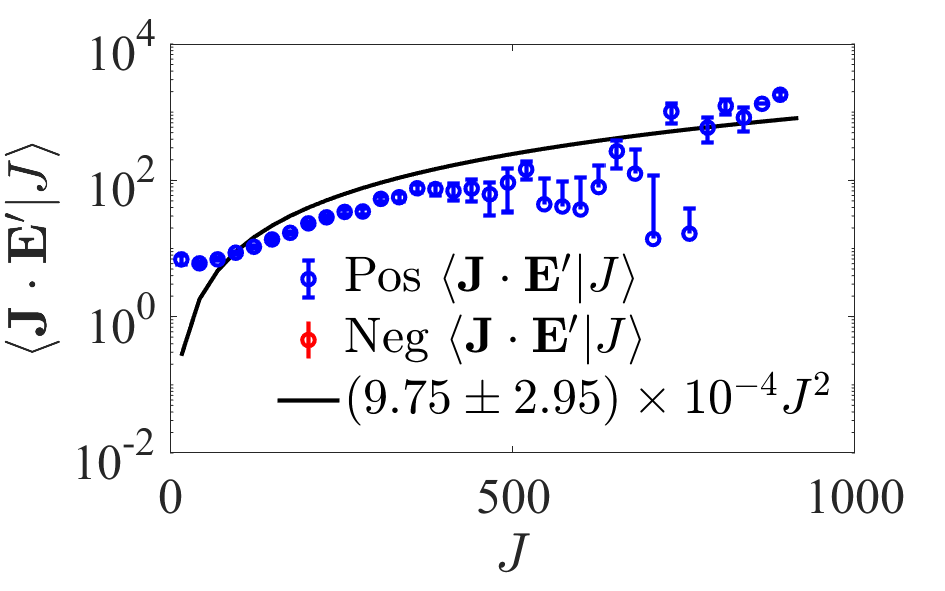}
    \includegraphics[width=0.32\textwidth]{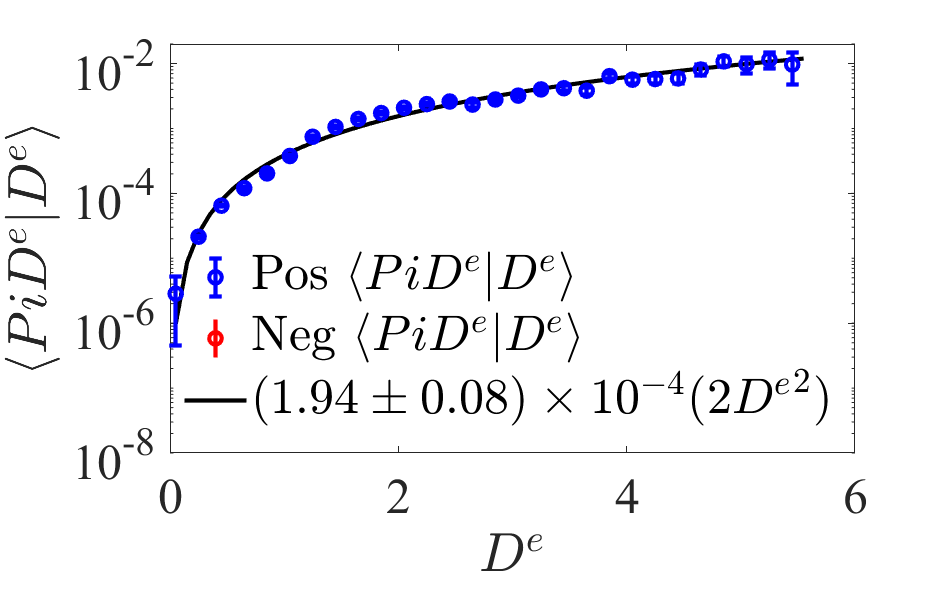}
    \includegraphics[width=0.32\textwidth]{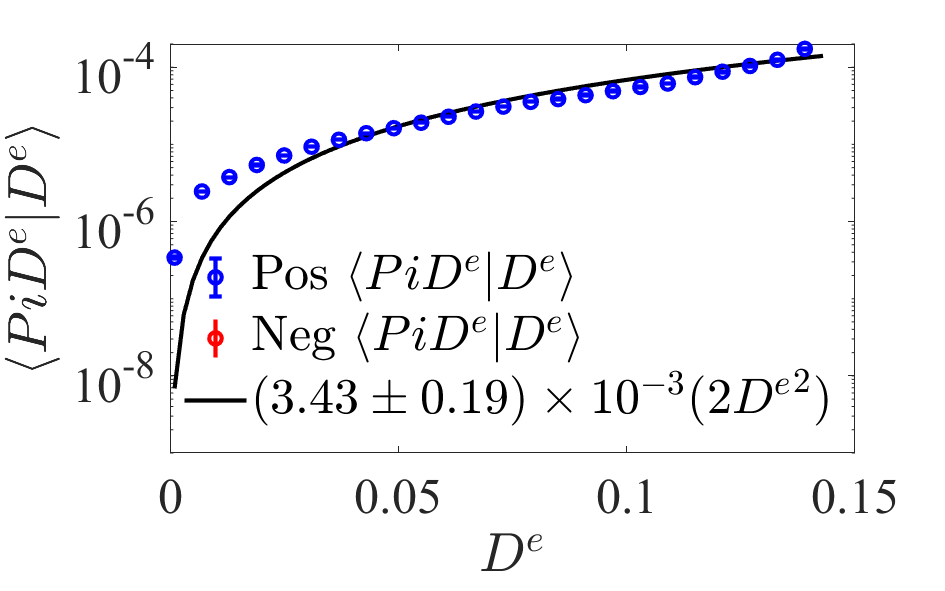}
    \includegraphics[width=0.32\textwidth]{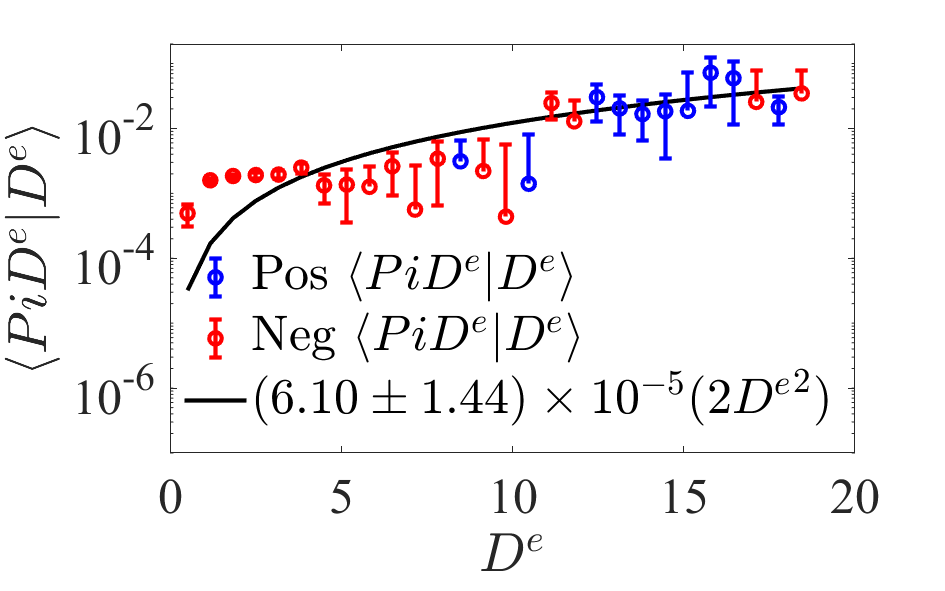}
    \includegraphics[width=0.32\textwidth]{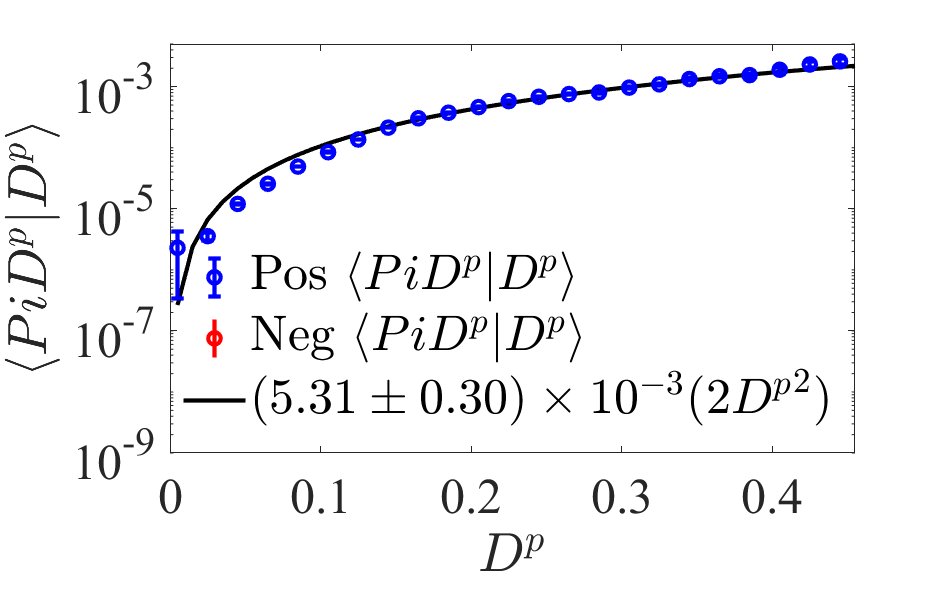}
    \includegraphics[width=0.32\textwidth]{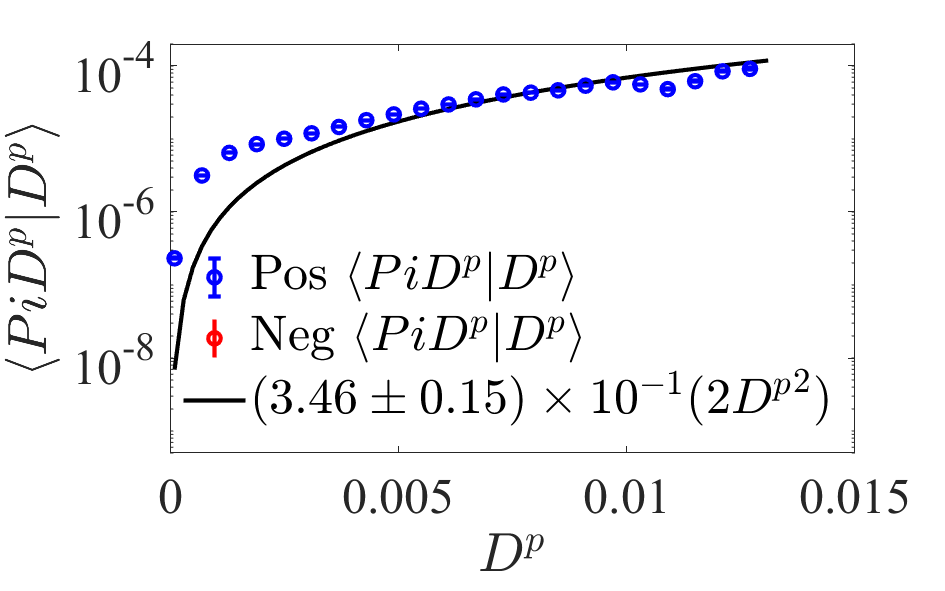}
    \includegraphics[width=0.32\textwidth]{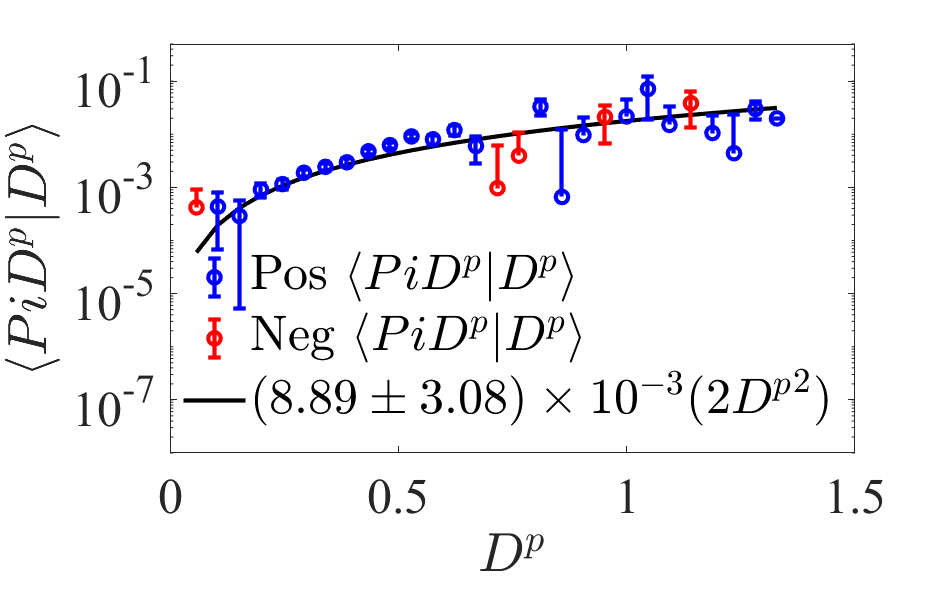}
    \caption{Conditional average of (top) the electromagnetic work ${\vct J} \cdot {\vct E}'$ with respect to the current density magnitude $J$, 
    and (middle and bottom) $ \PiD^{(\alpha)}$ (= $-\Pi_{ij}^{(\alpha)} \, D_{ij}^{(\alpha)} $) 
    with respect to the traceless velocity strain rate $ D^{(\alpha)} = \sqrt{D_{ij}^{(\alpha)} \, D_{ij}^{(\alpha)}} $. The error bars are computed from the standard deviation in each bin. The coefficients from least-square fitting, within $95\%$ confidence interval, are also shown.}
    \label{fig:vis-res}
\end{figure*}

To determine the values of the effective diffusion coefficients,
we employ a method based on the recent work of 
\citet{BandyopadhyayEA2023-collisionallike}.
The basic procedure, for the case of resistivity determination,
is to compute 
$\langle \vct{J} \cdot \vct{E}^\prime |J \rangle$,
which is the average
of the electric work in the electron fluid frame conditioned on the (local) current magnitude 
    $J = |\vct{J} |$,
and investigate its dependence on $J$.
As was noted previously 
\citep{WanEA16,BandyopadhyayEA2023-collisionallike},
this conditional average
is found to follow a curve 
$\langle \vct{J} \cdot \vct{E}^\prime |J \rangle \sim J^2$
to a reasonable degree of accuracy, as shown here in the top row of 
Fig.~\ref{fig:vis-res}. 
The error bars are computed from the standard deviation in each bin.
Using this quadratic scaling agreement we evaluate the 
constants of proportionality for the two simulations and for the 
MMS data, thus providing 
an estimation of the effective resistivity for the respective cases.
The functional form of the trend is strongly similar to that of the collisional case, 
as given in Eq.~\eqref{eq:res_diss}. This accounts for the 
heuristic description of the result as ``collisional-like''. 
These values of ``effective resistivity'' $1/\sigma$, within $95\%$ confidence interval,
are shown in the legend of 
Fig.~\ref{fig:vis-res} and tabulated in 
Table~\ref{tab:vis-res} in the respective units. 

A similar procedure is followed for the 
the conditional average of the anisotropic part of the pressure-strain
interaction, $ \PiD^{(\alpha)}$ (= $-\Pi_{ij}^{(\alpha)} \, D_{ij}^{(\alpha)} $), 
which represents the incompressive contribution
to the rate of production of thermal energy
\citep{Braginskii65,ChiuderiVelli,YangEA22}.
This is done separately for electrons and protons.
Specifically, we compute the average  
of electron $ \PiD^{e}$ conditioned on 
 $ D^e \equiv \sqrt{ D^e_{ij}D^e_{ji} } $, that is
 $ \langle - \Pi_{ij}^e D_{ij}^e| D^e \rangle$.
Recall that the traceless strain rate tensor for the electron fluid velocity is 
 $ D^e_{ij} = \frac{1}{2}  (\partial_i u^e_j + \partial _j u^e_i) - \frac13 \delta_{ij} \nabla \cdot \vct{u}^e $.
The general trend is quite consistent with the collisional 
scaling in Eq.~\eqref{eq:vis_diss}, as shown in the second row of Fig.~\ref{fig:vis-res}.
Thus the resulting approximation, that 
$\langle - \Pi_{ij}^e D_{ij}^e| D^e \rangle \propto (D^e)^2$, 
is indeed a collisional-like representation of the average results. 

The analysis for the proton case proceeds in direct analogy to the electron case, with the results shown in the third row of Fig.~\ref{fig:vis-res}. 
The conditional average of proton $ \PiD^{p}$ is also found to be well approximated by a fit to a collisional-like scaling, as described in Eq.~\eqref{eq:vis_diss}. 
That is, 
$\langle - \Pi_{ij}^p D_{ij}^p| D^p \rangle \propto (D^p)^2$, where $D^p_{ij}$ is the traceless strain rate tensor for the proton fluid velocity
and 
  $ D^p \equiv \sqrt{ D^p_{ij}D^p_{ji} } $. 

One might notice that, unlike the positive definite collisional dissipation in Eqs.~\eqref{eq:vis_diss} and \eqref{eq:res_diss}, $\langle -\Pi_{ij}^{(\alpha)} \, D_{ij}^{(\alpha)} |D^{(\alpha)} \rangle$ is sign-indefinite, especially for the MMS data. Here we presume the existence of uniform viscosity and resistivity, without taking into account the sign effect. That is, for any negative conditional averages, we are taking their absolute values to fit the collisional scalings. A more careful and refined treatment of the sign effect is deferred to a future study.

All of the above results, including computations
of the effective kinematic viscosity ($\nu=\mu/\rho$) and magnetic diffusivity ($\eta=1/(\sigma \mu_0)$) are shown in Table~\ref{tab:vis-res}.
Note that the diffusion coefficients from P3D and VPIC are expressed in the respective code units.
%see Table.~\ref{tab:2.5d-3d-units}. 
%To be consistent with the MMS interval, the code units in P3D and VPIC are 
%evaluated in SI units using 
%using the measured values obtained for the MMS interval. 
To facilitate a direct comparison of the simulation numbers with MMS, we use the plasma properties measured for the MMS interval to convert the diffusion coefficients from code units to SI. That is, to compute the units in P3D, we need to use the proton inertial length $\di$, proton cyclotron frequency $\omega_{cp}$, Alfv\'en speed $V_A$, mean electron number density $\langle n_e \rangle$ measured over the MMS interval, and the real-life values of proton mass 
  $ m_\text{p} = 1.67 \times 10^{-27} $\,kg 
and proton charge 
  $ e = 1.6 \times 10^{-19}$\,C. 
Similarly, to compute the VPIC units, we need to use the electron inertial length $\de$, electron plasma frequency $\omega_{pe}$, mean electron number density $\langle n_e \rangle$ from the MMS interval, and the real-life values of electron mass $m_\text{e}$, proton charge $e$, and speed of light $c$. Finally, the effective kinematic viscosity and effective magnetic diffusivity from the two PIC simulations and the MMS interval are all expressed in SI units, m$^2$/s; see Table~\ref{tab:vis-res-SIunits}. Note that the viscosity and diffusivity are widely distributed for different datasets, reflecting the physical difference between the simulations and MMS data.

\begin{table*}
%    \large
    \centering
    \begin{tabular}{l|l|l|l}
    \hline
	    Variables  & 2.5D P3D & 3D VPIC & MMS\\
     \hline
	    $\frac{1}{\sigma}$ & $(5.57\pm1.09)\times 10^{-3} ~[\rm \frac{m_\text{p}\omega_{cp}}{n_r e^2}]$ & $(3.80\pm0.11)\times 10^{-3} ~[\rm \frac{m_\text{e}\omega_{pe}}{n_r e^2}]$ & $(9.75\pm2.95)\times 10^{-4} ~[\rm \frac{mV \cdot m}{nA}]$\\
     \hline
        $\eta=\frac{1}{\sigma \mu_0}$  & $(4.43\pm0.87)\times 10^{3} ~[\rm \frac{m_\text{p}\omega_{cp}}{n_r e^2} \frac{m}{H}]$ & $(3.02\pm0.09)\times 10^{3} ~[\rm \frac{m_\text{e}\omega_{pe}}{n_r e^2} \frac{m}{H}]$ & $(7.76\pm2.35)\times 10^{8} ~[\rm m^2/s]$ \\
      \hline
 $\mu_e$ & $(1.94\pm0.08)\times 10^{-4} ~[\rm m_\text{p} n_r V_{Ar} \di]$ & $(3.43\pm0.19)\times 10^{-3} ~[\rm m_\text{e} n_r c \de]$ & $(6.10\pm1.44)\times 10^{-5} ~[\rm nPa \cdot s]$\\
      \hline
        $\nu_e=\frac{\mu_e}{\rho_e}$  & $(4.85\pm0.20)\times 10^{-3} ~[\rm V_{Ar} \di]$ & $(3.43\pm0.19)\times 10^{-3} ~[\rm c \de]$ & $(2.69\pm0.64)\times 10^{9} ~[\rm m^2/s]$\\
      \hline
        $\mu_p$ & $(5.31\pm0.30)\times 10^{-3} ~[\rm m_\text{p} n_r V_{Ar} \di]$ & $(3.46\pm0.15)\times 10^{-1} ~[\rm m_\text{e} n_r c \de]$ & $(8.89\pm3.08)\times 10^{-3} ~[\rm nPa \cdot s]$\\
      \hline
      $\nu_p=\frac{\mu_p}{\rho_p}$  & $(5.31\pm0.30)\times 10^{-3} ~[\rm V_{Ar} \di]$ & $(6.92\pm0.30)\times 10^{-3} ~[\rm c \de]$ & $(2.33\pm0.81)\times 10^{8} ~[\rm m^2/s]$ \\
      \hline
    \end{tabular}
    \caption{Effective electrical resistivity $1/\sigma$ and 
    effective dynamic viscosity $\mu$ within $95\%$ confidence interval from least-square fitting in Fig.~\ref{fig:vis-res}, and the corresponding effective kinematic viscosity $\nu=\mu/\rho$ and effective magnetic diffusivity $\eta=1/(\sigma \mu_0)$. The units are shown enclosed in square brackets and are those that apply to the specific code or data interval.}
    \label{tab:vis-res}
\end{table*}

\begin{table*}
%    \large
    \centering
    \begin{tabular}{l|l|l|l}
    \hline
	    Variables  & 2.5D P3D & 3D VPIC & MMS\\
     \hline
        $\eta~[\rm m^2/s]$  & $(2.44\pm0.48)\times 10^{7}$ & $(1.21\pm0.04)\times 10^{9} $ & $(7.76\pm2.35)\times 10^{8} $ \\
      \hline
        $\nu_e~[\rm m^2/s]$  & $(2.27\pm0.09)\times 10^{7}$ & $(1.13\pm0.06)\times 10^{9}$ & $(2.69\pm0.64)\times 10^{9}$\\
      \hline
        $\nu_p~[\rm m^2/s]$  & $(2.49\pm0.14)\times 10^{7}$ & $(2.28\pm0.10)\times 10^{9} $ & $(2.33\pm0.81)\times 10^{8}$ \\
      \hline
    \end{tabular}
    \caption{Effective kinematic viscosity $\nu$ and 
    effective magnetic diffusivity $\eta$ from Table~\ref{tab:vis-res} re-expressed in SI units: m$^2$/s.}
    \label{tab:vis-res-SIunits}
    \end{table*}
%------------------------------------------------------------------

    \subsection{Empirical determination of Reynolds numbers}
    \label{sec:Rey}
    
Given a diffusivity, a general prescription to obtain a Reynolds number ($ \Rey $) is to assemble 
\begin{equation}
  \Rey = \frac{ {\rm speed } \times {\rm length}}
{\rm diffusivity},
\label{eq:Re-formal}
\end{equation}
where the speed and length are those characteristics 
of the turbulence. %They may be determined or estimated in at least several ways.
The results in the previous section
make it possible to compute \emph{effective Reynolds numbers} $\Rey$ as described in 
Eq.~\eqref{eq:Re-formal}, 
since we now have
quantitative values for the 
(effective) diffusivities
$\eta$, $\nu_e$ and $\nu_p$.  
%One option that presents immediately 
%is to compute \emph{effective Reynolds numbers}
%using the empirically determined 
%viscosity
%and diffusivity.
Choosing the correlation scales for the species velocities, $\lambda_{c,\alpha}$ 
($\alpha=e, p$ for electrons and protons, respectively),  and magnetic field, $\lambda_{c,b}$,
as the characteristic lengths,
we may write separate
effective Reynolds numbers for the electron and protons, $\Rey_{c,\alpha}$, 
and 
an effective magnetic Reynolds number, $\Rey_{c,b}$, 
as
\begin{align}
  \Rey_{c,\alpha} 
      &=
      \frac{u_{\alpha} \lambda_{c,\alpha}}{\nu_{\alpha}},
 \label{eq:Re}\\
  \Rey_{c,b}
      &=
      \frac{u \lambda_{c,b}}{\eta}.
\label{eq:Reb}
\end{align}
Here, $u_{\alpha}$ are the characteristic fluctuation speeds for each species.
For the magnetic Reynolds number, the characteristic speed is denoted $u$ and there is some flexibility in deciding what value to use for it.

The required values 
of \emph{correlation scale} can be obtained as follows: 
The scale-dependent auto-correlation function is defined as
\begin{equation}
R(\vct{r})=\frac{\langle \vct{F}(\vct{x}+\vct{r}) \cdot \vct{F}(\vct{x}) \rangle}{\langle \vct{F}(\vct{x}) \cdot \vct{F}(\vct{x}) \rangle},
\label{eq:cor}
\end{equation}
where $\vct{F}$ can be either the fluctuation velocity or magnetic field.
Note that $R(\vct{r})$ is a function of lag vector $ \vct{r} = (r_x, r_y, r_z) $. Upon averaging over directions, $R(\vct{r})$ only depends on lag length $r = |\vct{r}|$, 
and $R(r)$ denotes the omnidirectional form of the auto-correlation function.
Based on computation 
of the auto-correlation function, 
the correlation scale $\lambda_c$ can be defined in several ways. Here we employ the 
so-called 
``e-folding'' method, 
\begin{equation}
    R(\lambda_c) = 1 / \e,
    \label{eq:e-folding}
\end{equation}
where the correlation scale is computed as the scale where the auto-correlation function drops to $1 / \e$,
on the basis that $\e^{-r/\lambda_c}$
is an adequate approximation for $R(r)$ \citep{MattEA99-swh,SmithEA18}.

\begin{figure*}
    \centering
    2.5D P3D \hspace{18em} 3D VPIC \hspace{18em} MMS\\
    \includegraphics[width=0.32\textwidth]{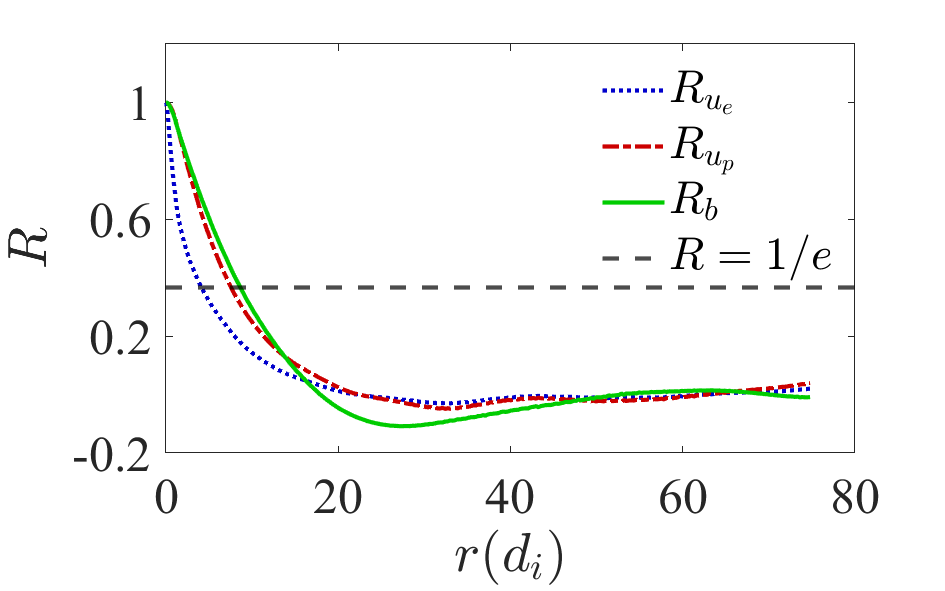}
    \includegraphics[width=0.32\textwidth]{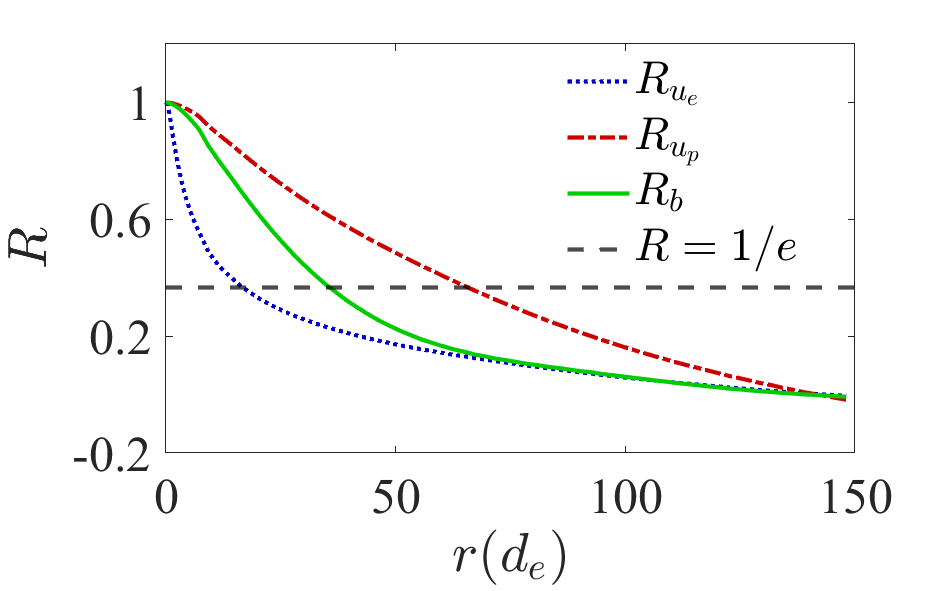}
    \includegraphics[width=0.32\textwidth]{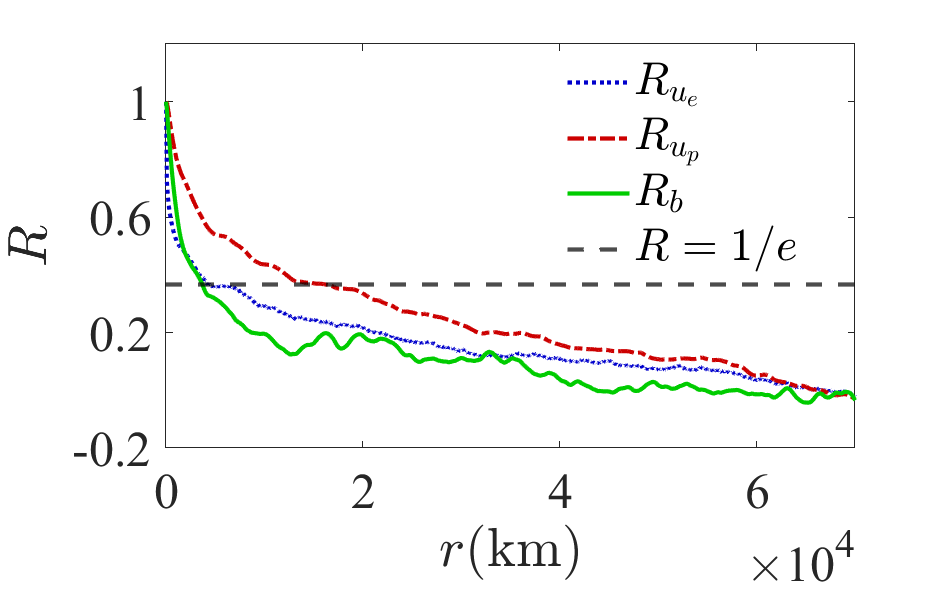}
    \caption{Correlation functions for the electron velocity, proton velocity, and magnetic field for the two PIC simulations and the MMS interval.}
    \label{fig:corr-fun}
\end{figure*}

Figure~\ref{fig:corr-fun} 
shows the results of our correlation analysis of simulation data and MMS observations.
These results are employed to extract correlation lengths.
%% and thence to compute effective Reynolds numbers. 
The average bulk speed in this MMS interval is approximately $V_{SW}=230\,\rm km\,\rm s^{-1} $, which is used to convert temporal scales to spatial scales for the MMS data.
Characteristic fluctuation
speeds and correlation scales 
for these three datasets are 
recorded in Table~\ref{tab:corr-scale-Re}.
  %% for the 2.5D and 3D PIC simulations and for the MMS data.
As the two codes use different normalizations, 
  Table~\ref{tab:corr-scale-Re}
also indicates the relevant normalizing quantities, or units, in square brackets. 
Together with the (effective) kinematic viscosities and magnetic diffusivities listed in Table~\ref{tab:vis-res}, these are combined in accordance with Eqs.~\eqref{eq:Re} and \eqref{eq:Reb}
to compute the three corresponding effective 
Reynolds numbers, 
shown also in Table~\ref{tab:corr-scale-Re}.
Notably, for the simulation cases 
the three Reynolds numbers are all rather similar, whereas for the MMS interval there are sizable differences, which could be attributed to the uncertainties when computing the correlation length.
\begin{table}%% [H]
%    \large
    \centering
    \begin{tabular}{c|l|l|c}
    \hline
	    Variables  & 2.5D P3D & 3D VPIC & MMS\\
     \hline
	    $u_e=\sqrt{\langle \vct{u}_e^2\rangle}$ & $0.30 ~[\rm V_{Ar}]$ & $0.055 ~[\rm c]$ & $232.8~[\rm km/s]$\\
     \hline
        $u_p=\sqrt{\langle \vct{u}_p^2\rangle}$ & $0.22 ~[\rm V_{Ar}]$ & $0.032 ~[\rm c] $ & $242.1~[\rm km/s]$\\
      \hline
        $\lambda_{c,e}$ & $4.1 ~[\rm \di]$ & $17 ~[\rm \de]$ & $4380~[\rm km]$\\
      \hline
        $\lambda_{c,p}$ & $7.5 ~[\rm \di]$ & $65 ~[\rm \de]$ & $15990~[\rm km]$\\
      \hline
        $\lambda_{c,b}$ & $8.5 ~[\rm \di]$ & $36 ~[\rm \de]$ & $3690~[\rm km]$\\
      \hline
       $\Rey_{c,e}=u_e\lambda_{c,e}/\nu_e$ & $250$ & $270$ & $380$\\
     \hline
        $\Rey_{c,p}=u_p \lambda_{c,p}/\nu_p$ & $320$ & $300$ & $16610$\\
      \hline
	    $\Rey_{c,b}=u_p \lambda_{c,b}/\eta$ & 370 &  310 & 1150 \\
      \hline
    \end{tabular}
    \caption{Characteristic fluctuation speeds $u$, correlation scales $\lambda_{c}$, and effective large-scale Reynolds numbers $\Rey_{c}$. Additional $e$, $p$, and $b$ subscripts indicate the quantity for the electrons, protons, or magnetic field, respectively.
    Here the characteristic fluctuation speed for protons, $u_p$, is used to compute the effective magnetic Reynolds number.}
    \label{tab:corr-scale-Re}
\end{table}

    \section{Discussion and Conclusions}
    \label{sec:disc}

This paper 
elaborates and extends the previous work by 
\citet{BandyopadhyayEA2023-collisionallike}
in which the initial analysis of conditional averages
was presented, indicating that a collision-like dissipation may be present in collisionless plasma, 
as suggested by consistency of the data with  
Eqs.~\eqref{eq:vis-scaling}
and \eqref{eq:res-scaling}.
Here we have quantitatively examined these approximate relations and determined
effective viscosities for electrons and protons as well as an effective resistivity. This was carried out separately for two plasma kinetic (PIC) simulations, one 2.5D and one 3D, 
and for a sample of magnetosheath turbulence  data recorded by the MMS mission. 
Having determined effective diffusion coefficients, 
and using measured fluctuation speeds and correlation scales, the 
assembly of effective Reynolds numbers follows directly. 

From the effective large-scale Reynolds number,   $\Rey_c$,
relationships involving the plasma equivalent of the 
Kolmogorov \emph{dissipation scale}, $\lambda_D$, may also be formulated,
    \begin{equation}
        \frac{\lambda_c}{\lambda_{D}}=C_{\epsilon}^{1/4} \Rey_c^{3/4},
         \label{eq:diss-scale}
    \end{equation}
where $C_{\epsilon}$ is the dimensionless dissipation rate.
This relation is formulated based
on the classic development  in hydrodynamic turbulence theory \citep{Kol41a,Kol41c,BatchelorTHT}, 
and may be used in several ways. 
One might substitute measured correlation scales $\lambda_c$ and effective Reynolds numbers $\Rey_c$ into the formula to extract an estimate of the dissipation scale $\lambda_D$. 
Alternatively one might assume, as has been done previously,  that the dissipation scale in a plasma such as solar wind, corresponds to the upper end of the inertial range. Then if the value of $\lambda_D$ is 
taken to be, for example 
the ion (or electron) inertial length $ \di$ (or $ \de$), 
    Eq.~\eqref{eq:diss-scale} 
may be construed as providing another alternative 
estimate of (effective) Reynolds number. 
There are also other approaches for estimating the dissipation scale.
For example,  
if the cascade rate $\epsilon$ is known and an effective viscosity is available,
hydrodynamic turbulence theory provides the Kolmogorov-style estimate  
$\lambda_{D} = \left( {\nu^3}/{\epsilon} \right)^{1/4}$.

The Reynolds numbers determined here 
are roughly consistent with reasonable estimates of the corresponding 
dissipation scales, through the formulation given by Eq.~\eqref{eq:diss-scale}.
%In particular, 
%there is a rough correspondence between the 
%determination of the Reynolds numbers and 
%reasonable estimate of dissipation scale, through the 
%alternative formulation given by Eq.~\eqref{eq:diss-scale}.
For example, substituting the
proton Reynolds number  $\Rey_{c,p}=320$ 
for the 2.5D simulations into 
Eq.~\eqref{eq:diss-scale} and
using a value $C_\epsilon = 0.5$ (as in, e.g., \cite{LinkmannEA17-Ceps,BandyopadhyayEA18-Ceps,Li2023anomaly})
and the measured correlation scale $\lambda_{c,p} = 7.5 \, \di$,
the relation Eq.~\eqref{eq:diss-scale} gives $\lambda_D \sim 0.1 \,\di$, which is not an unreasonable estimation.  
For the MMS data, the same line of analysis leads to the estimate
$\lambda_D \sim 13 $\,km.  This too is a reasonable estimate for 
the dissipation scale in the magnetosheath, where for this interval the value of 
$\di$ is 48\,km (see Table~\ref{tab:MMS}).
In fact,
there are several other ways to combine the above values of Reynolds numbers and 
measured parameters to examine consistency with traditional estimates. %such as 
%Eq.~\eqref{eq:diss-scale} including the latter with $\lambda_D$ estimated in a more or less traditional way, as being 
%of order $ \di$. 
All the combinations
we have tried provide reasonable answers,
such as values of $\lambda_D$ 
that are deemed reasonable given the 
findings from simulations and other observations
that spectral steepening usually occurs near $ \di $.
However, no firm guidance is available providing a more detailed 
picture of a scale at which 
electron and proton dissipation 
become dominant or 
``turn on'' relative to each other.
See, for example, \cite{YangEA22}.

Another interesting aspect of the present results is the size of the (effective) magnetic 
Prandtl number $Pm$, generally defined as the ratio of kinematic viscosity $\nu$ to magnetic diffusivity $\eta$.  
Here, examining the 
values of the effective magnetic diffusivity and two viscosities stated in 
    Table~\ref{tab:vis-res-SIunits},
we see that the magnetic Prandtl number estimates from both simulation 
results are near unity.
For the MMS data, the value is 
only moderately away from unity. 
The significance of this is that when 
 $Pm$
%% the magnetic Prandtl number 
greatly 
differs from unity, different regimes 
of MHD scale behavior become possible 
  \citep{ChoEA02,PontyEA05,SahooEA11}. 
In particular, in such cases 
the inertial ranges in magnetic field
and velocity field can develop very different bandwidths. 
A value of $Pm$ near unity is consistent with the usual finding of ``Alfv\'enic'' turbulence in which there is order 
one equipartition 
of magnetic and velocity field energy in their respective inertial ranges over 
very similar ranges of wavenumber. 

We remark that although the classical collisional diffusivity must obviously be absent in collisionless plasmas, a number of previous 
works have nonetheless attempted to write approximate expressions for effective diffusion coefficients in collisionless 
plasma. Possible candidates that act as effective collisions include wave-particle interactions \citep{graham2022direct}, pitch angle scattering \citep{earl1988cosmic,zank2014pickup}, stochastic field line effects, and other kinetic mechanisms.
In particular viscosity and resistivity
have been estimated based on 
various approximations; for the present, we leave aside the estimation of other transport coefficients such as heat conduction \citep{Hollweg76,RiquelmeEA16}. 
Resistivity is often estimated using terms in the generalized Ohm's law in terms of fluid quantities \citep{graham2022direct,SelviEA23}, 
or in the case of hyper-resistivity, by consideration of contributions from anomalous electron viscosity \citep{Strauss86-hyper}. 
A collisional-like viscosity is already present in earlier studies, such as gyroviscosity \citep{smolyakov1998gyroviscous}, cosmic-ray viscosity \citep{earl1988cosmic}, and plasma viscosity \citep{kaufman1960plasma}.
Viscous effects are often estimated by consideration of the MHD-scale cascade and its implications for
pressure anisotropies 
\citep{QuataertGruzinov99,SharmaEA07,Verma2019EPJB}.
Considerations of pressure anisotropy
  \citep{KasperEA02,MatteiniEA07}
and linear instabilities that drive it, can be employed to develop theories for effective viscosity. 
This may be particularly 
effective when combined with 
exact results from Vlasov--Maxwell theory, such as dissipation through the pressure-strain interaction
   \citep{YangEA17-PoP}.
Such considerations 
have motivated 
more elaborate approximate 
models for effective viscosity
based on pressure anisotropy in the simplified CGL model
\citep{squire2023pressure,ArzamasskiyEA23}. 
Valuable insights are obtained from models of this type, especially  with regard 
to extrapolations to extreme values 
of plasma $\beta$ that can be relevant
to astrophysical systems 
\citep{KawazuraEA19,Howes10,RoyEA22-QiQe}. 

Finally, we recall that several 
additional relationships may be 
adapted from classical turbulence 
theory to provide 
alternative estimates for Reynolds numbers and diffusivities.
One possibility is to base 
measurements on the 
Taylor microscale $\lambda_T$,
which can be related directly to 
the second derivative of the 
auto-correlation function evaluated at the origin
 \cite[e.g.,][]{BatchelorTHT,Pope}. 
Up to order unity constants, 
  $ \lambda_T = [-R''(0)]^{-1/2} $,
where we have in mind that $R(r)$ 
is the direction averaged correlation function of, say, the magnetic field, as in Eq.~\eqref{eq:cor}.
Then one can show that an estimate of the effective Reynolds number can be written as 
  $ \Rey = (\lambda_c / \lambda_T)^2/C_{\epsilon} $.
This quantity is measurable when high-resolution data is available, and may be further developed into an estimate of the effective viscosity, as shown by 
  \citet{BandyopadhyayEA20-Taylor}.
The above relationships should be viewed as semi-empirical
and, while motivated by theory, 
should not be treated as exact in any sense, since the underlying theories are mainly hydrodynamic (and collisional) 
and usually founded on simple assumptions of rotational symmetry or incompressibility.

Based on the above results and 
given the unique nature of the analysis developed so far, we conclude that the present approach to quantifying 
collisional-like dissipation
in 
collisionless plasma turbulence 
warrants further investigation. Already we have seen herein
that examination of 
conditional averages of pressure-strain interaction and electric work, which themselves are exact statements of energy conversion rates, provides a basis for finding effective diffusion coefficients. 
With apparently 
reasonable values of (effective) Reynolds numbers, viscosities, and resistivities 
in hand, the door is opened to 
examining a class of relationships 
that may help bring 
turbulence theory concepts into 
greater contact with turbulent plasma,
as we have described above.
There is clearly much more
to do in the complex subject of 
plasma turbulence, 
and the present work offers
a small step in 
a possibly useful 
direction. 
 
\section*{Acknowledgements}
\begin{CJK*}{UTF8}{gbsn}
This research is supported in part by the 
MMS Theory and Modeling grant 80NSSC19K0565 at Delaware,
and by NSFDOE grant PHYS-2108834 at Delaware,
and NASA Heliospheric GI Grant No.\ 80NSSC21K0739 
and NASA Grant No.\ 80NSSC21K1458 at Princeton University,
and a subcontract SUB0000317 to Delaware.
M. A. S. acknowledges support from NASA LWS grant 80NSSC20K0198.
We would like to acknowledge high-performance computing support from Cheyenne (doi:10.5065/D6RX99HX) provided by NCAR's Computational and Information Systems Laboratory, sponsored by the National Science Foundation.
This research also used resources of the National Energy Research Scientific Computing Center, which is supported by the Office of Science of the U.S. Department of Energy under Contract No. DE-AC02-05CH11231.
We are particularly 
grateful to 
Sylvie Yang Jin (金浠言)
for cooperation and assisting the lead author in 
completing this 
research. 
\end{CJK*}

\section*{Data Availability}
This study used Level 2 FPI and FIELDS data according to the guidelines set forth by the MMS instrumentation team. The data that support the findings of this study are openly available in MMS SDC at \url{https://lasp.colorado.edu/MMS/sdc/}.
Other reasonable requests for sharing of the metadata regarding computational runs, custom simulation codes and documentation will generally be honored.
%-----------------------------------
%\bibliographystyle{mnras}
%\bibliography{A-Z,refs_YY,refs_WHM}

%\bsp	% typesetting comment
%\label{lastpage}
%\end{document}

 \newcommand{\BIBand} {and} %...... how 'and' appears in authors
  \newcommand{\boldVol}[1] {\textbf{#1}} %......................
  \providecommand{\SortNoop}[1]{} %.......Use as {\SortNoop{Aaa}}
  \providecommand{\sortnoop}[1]{} %..............................
  \newcommand{\stereo} {\emph{{S}{T}{E}{R}{E}{O}}} %................
  \newcommand{\alfven} {{A}lfv{\'e}n\ } %...........................
  \newcommand{\alfvenic} {{A}lfv{\'e}nic\ } %.........................
  \newcommand{\Alfven} {{A}lfv{\'e}n\ } %...........................
  \newcommand{\Alfvenic} {{A}lfv{\'e}nic\ }

\bsp	% typesetting comment
\label{lastpage}
\end{document}